# Is There a Stable Bucky-diamond Structure for SiC Clusters?

*Ming Yu, C.S. Jayanthi, and S.Y. Wu*
*Department of Physics and Astronomy, University of Louisville, Louisville, KY 40292*

Abstract

We have carried out an extensive search for the SiC Bucky-diamond structure to confirm not only that a pair of Si and C atoms can form $sp^2$- as well as $sp^3$-type bonds but also that these two types of bonds can co-exist in the same SiC-based structure. The successful and surprising discovery of the $Si_nC_m$ Bucky-diamond structure at the specific composition of $n=68$ and $m=79$ is the result of the relaxation of the truncated bulk *3C*-SiC network to yield a $Si_nC_m$ cluster with $m+n=147$. It highlights the important role played by the composition in determining the structure and hence other properties of SiC-based nano-structures. We have also shed light on the mechanism behind the formation of the Bucky-diamond structure. The formation process is initiated by the induced bonds between pairs of surface carbon atoms of the initial configuration of the $Si_{68}C_{79}$ cluster obtained by truncating bulk *3C*-SiC network. This action then continuously incorporates atoms in the six outer shells of the $Si_{68}C_{79}$ clusters to form the 112-atom Fullerene shell through nearest neighbor Si-C interactions. Because the 35-atom inner core with five completely filled shells only interacts weakly with the Fullerene shell through the six atoms on its "surface", the diamond-like inner core is barely perturbed and is suspended inside the Fullerene shell. We have also suggested a likely route of synthesizing the SiC Bucky-diamond structure based on the result of our simulation.

**I. Introduction**

Silicon carbide based nanostructures have recently been explored extensively because of their promising technological applications [1-4]. These materials are promising electronic materials for devices and have the potential as photovoltaic materials for solar cells, electrode materials for Li-ion batteries, *etc*. Bulk SiC possesses extraordinary mechanical and physical properties [5-12]. It is a material with low density, high strength, high thermal conductivity, high index of refraction, low thermal expansion, and wide band gap. It is stable at high temperature and chemically inert [13-15]. These exceptional intrinsic properties make SiC-based nanostructures outstanding candidates as environmentally friendly light weight materials for energy, electronics, and drug-delivery applications.

It is well known that Si and C atoms bond through $sp^3$ type bonding in bulk SiC. But, in a recent theoretical study, we have shown that graphitic-like 2-D SiC sheets are stable, indicating the feasibility of a $sp^2$ type bonding between Si and C atoms [16]. The above facts imply that bonding between a Si atom and a C atom must be similar to that between two C atoms. It forms a $sp^3$ bonding when in the (3-D) bulk and forms a $sp^2$ bonding when in the (2-D) sheet, opening up the possibility of finding the same sorts or types of nanostructures as those found in carbon (*e.g.*, compact clusters, caged clusters, fullerenes, nano-wires (NWs), nanotubes (NTs), *etc.*).

The most stringent test for validating the similarity of the bonding nature between a pair of C atoms and that between a Si atom and a C atom is to determine whether there could be a stable Bucky-diamond structure for SiC clusters. In the Bucky-diamond $C_{147}$, there is a diamond core of 35 C atoms bonded by $sp^3$ bonds in a tetrahedral network, with a shell of 112 C atoms bonded by $sp^2$ bonds in a Fullerene structure [17-19]. There is only weak interaction between the inner core and the outer shell as evidenced by the fact that the nearest neighbor distances between the surface atoms in the inner core and atoms in the outer shell range from 1.60 to 3.34 Å [17, 18], far longer than either the $sp^2$ bond length in graphene sheet (1.42 Å) or the $sp^3$ bond length in diamond (1.52 Å). Hence the Bucky-diamond structure must be a structure where $sp^2$ and $sp^3$ bonds co-exist in the same structure, leading to a $sp^3$ bonded diamond inner core weakly coupled to a $sp^2$ bonded Fullerene shell.

Since a Si atom forms the $sp^2$ bond with a C atom in the 2-D graphitic-like sheet and a $sp^3$ bond in the bulk, there is then the likelihood that a SiC Bucky-diamond structure may exist. If so, the existence of this unusual structure of Si/C-based system should convincingly demonstrate the equivalence of the bonding nature between a Si atom and a C atom in Si/C-based systems and that between a pair of C atoms in C-based systems. It also implies that, similar to the case of carbon nanostructures, SiC-based nano-structures could exist in multiple forms (*i.e.* polymorphic structures). Furthermore, in the case of SiC-based systems, there is one more factor, the composition, that could be manipulated. The manipulation of this extra factor could provide additional routes to explore the geometrical and functional possibilities for stable SiC-based nano-structures. Thus, the study of the likelihood of the existence of SiC Bucky-diamond structure is not only academically interesting, but also could have implications for innovative device-related applications.

In this work, we studied the possible existence of the Bucky-diamond structure of $Si_nC_m$ clusters for $n+m=147$ using the SCED-LCAO based molecular dynamics (MD) method [20], (where SCED refers to the self-consistent and environment-dependent Hamiltonian and LCAO to linear combination of atomic orbitals). Specifically, we investigated the role played by the composition in possibly achieving a stable Bucky-diamond structure for $Si_nC_m$ clusters in terms of the interplay among the bonding nature, the bond length, and the bond strength between Si-Si, C-C, and Si-C atoms. We were able to determine that a stable Bucky-diamond structure indeed exists for the $Si_nC_m$ cluster with a specific composition ($n=68$ and $m=79$). We will demonstrate that the interplay among the above mentioned three factors, coupled with the suppression of the dangling bonds for the surface atoms, facilitates the formation of the stable Bucky-diamond structure of the $Si_nC_m$ cluster for a specific composition corresponding to $n=68$ and $m=79$.

The structural optimization of a SiC cluster of 147 atoms can, in principle, be studied through DFT calculations [21]. However, with no *a priori* knowledge of the composition for a plausible Bucky-diamond $Si_nC_m$ structure, an extensive search of different possible combinations of compositions is necessary. Such a search using the DFT-based method will be computationally expensive and thus not efficient for this particular problem. Therefore, we have used the molecular dynamics (MD) scheme based on the SCED-LCAO Hamiltonian (a semi-empirical Hamiltonian) to conduct the search for the SiC Bucky diamond structure. This semi-empirical Hamiltonian, that includes environment-dependent electron-ion interactions and evaluates charge re-distributions self-consistently, has predicted correctly the surface structures of silicon including Si(111)-7x7, low-dimensional structures of carbon (fullerenes, bucky-diamonds, carbon nanotubes, *etc.*), SiC graphene-like and graphitic structures, and bulk structures of Si, C, and SiC [17,18,20,22]. The predictions from the SCED-LCAO method have been extensively tested against the results from DFT based methods (using VASP). Such validations will also be conducted for the final relaxed structures of $Si_nC_m$.

In the present work, an extensive search for a plausible $Si_nC_m$ Bucky-diamond structure (with $n+m = 147$) was conducted by constructing various possible initial structures that included: (i) $Si_{35}C_{112}$ with all atoms arranged in a diamond-like network, (ii) $Si_{35}C_{112}$ with 35 Si atoms in the core arranged in a diamond-like network and the rest 112 C atoms in a Fullerene-like shell, (iii) spherical truncation of *3C*-SiC bulk such that the resulting $Si_nC_m$ network contained 147 atoms with $n = 112, 79, 73, 68, 35$ and $m = 35, 68, 74, 79, 112$, respectively with $n$ Si atoms in the interior core and $m$ C atoms on the exterior shells (Si core/C shell), a vice-versa arrangement, namely $m$ C atoms in the interior and $n$ Si atoms on the exterior (C core/Si shell), a segregated arrangement of $n$ Si and $m$ C atoms in the network, and finally an alternating arrangements of Si and C atoms as found in the bulk *3C*-SiC bulk, corresponding to $n= 79$ and $m=69$ as well as $n =68$ and $m=79$ (*i.e.* with Si as the central atom or C as the central atom). Section II will describe how our chosen intuitive initial structures given in (i) and (ii) failed to yield the Bucky diamond $Si_nC_m$ structures for $n+m =147$. Therefore, we considered initial structures outlined in (iii) where both compositions and distribution of Si and C atoms in the *3C*-SiC network are manipulated. This later study led to the discovery of the Bucky diamond $Si_nC_m$ structure for the magic composition $n=68$ and $m=79$. The formation route of the Bucky diamond $Si_{68}C_{79}$ from its initial diamond-like *3C*-SiC network structure is explained in section IV. The conversion of Bucky diamond $Si_{68}C_{79}$ to a cage structure of the same composition upon annealing and quenching is discussed in section V. The concluding remarks and the prospects for future extensions are given in section VI.

**II. Towards a Search for the Bucky Diamond $Si_{35}C_{112}$ Structure: Initial Structures Based on Diamond-like Tetrahedral Network and Carbon Bucky Diamond Structure**

It is known that carbon exhibits Bucky-diamond structure [17-19]. Specifically, for $C_{147}$ it is known that 35 interior atoms exhibiting $sp^3$ bonding form the core in the diamond network while 112 exterior carbon atoms with $sp^2$ bonding form a Fullerene-like shell. Since ordinarily Si atoms prefer $sp^3$ bonding while carbon atoms in a shell prefer the $sp^2$ bonding, it then seems prudent to "guess" that a plausible Bucky-diamond $Si_nC_m$ could be formed with a $Si_{35}$ core in the diamond network and a $C_{112}$ Fullerene-like shell. While this initial guess is intuitively very appealing, we decided to first test a more stringent scenario, namely, all atoms are arranged in a diamond-like tetrahedral network with 35 Si atoms filling up the first five inner shells and 112 carbon atoms occupying the six exterior shells. The bond length between any pair of atoms was chosen to be the same as the equilibrium bond length of bulk SiC (1.97Å).

This initial configuration of smaller core (compared to the bulk Si bond length) and larger shell (compared to the bulk C bond length) was designed to promote and expedite the relaxation to a stable equilibrium configuration. This initial structure was relaxed using the SCED-LCAO MD. The resulting process of approach to equilibrium, shown through the total energy/atom versus time curve, is given in Fig. 1a. It can be seen that the Si-Si bonds on the surface of the inner Si core start to break off right away as the "surface" Si atoms being pulled by the adjacent C atoms with stronger C-Si bonds towards the exterior. This process continues without interruption until there is no longer any inner core and all the Si atoms have migrated to the surface. The energy has also stabilized to within 0.002 eV/atom. The resulting equilibrated configuration is a complex cage-like structure with an open mouth. From the pair distribution of the equilibrated configuration shown in the inset of Fig. 1a, it can be seen that there are only nearest-neighbor C-C bonds and C-Si bonds, but hardly any nearest-neighbor Si-Si bond in the final "equilibrium" configuration. While the guessed trial initial configuration has not led to the Bucky-diamond structure for the $Si_{35}C_{112}$ cluster with this core/shell structure, it still demonstrates that a C-Si bond does favor the $sp^2$-bonding nature on the surface of a SiC-based structure. However, the result does not exhibit a structure where SiC $sp^2$ bonds co-exist with SiC $sp^3$ bonds.

Since the $Si_{35}C_{112}$ arranged in a diamond network did not lead to a Bucky-diamond structure, we decided to consider another initial structure favorably disposed to the formation of SiC Bucky-diamond structure. This new trial configuration is composed of 35 Si atoms filling up the first five shells of the tetrahedral network with a nearest-neighbor Si-Si bond length of 2.14 Å surrounded by a Fullerene-like shell of 112 C atoms with nearest-neighbor C-C distances in the range of 1.88-1.91 Å to accommodate the large core. The result is shown in Fig. 1b. It can be seen that the approach to equilibrium follows a very similar pattern as the case shown in Fig. 1a, leading to an almost equivalent final equilibrium configuration, *i.e.*, an open-mouthed cage structure.

The message conveyed by this study can be summarized as follows: (1) the cluster $Si_{35}C_{112}$ with its composition deduced from intuition will not lead to the stable $Si_nC_m$ Bucky-diamond structure; (2) a Si atom on the surface of a Si/C-based system may not have dangling bonds as long as it forms Si-C bonds with its carbon nearest-neighbors, another example of SiC bond being of $sp^2$ type on the surface rather than of $sp^3$ characteristics as in the bulk; (3) the likelihood of the existence of a stable $Si_nC_m$ Bucky-diamond structure must depend on a delicate balance of factors including the strength and the bonding nature of the Si-C bond, the C-C bond, and the Si-Si bond; the bond length of these three bonds; the composition of the $Si_nC_m$ ($m+n=147$) cluster. Therefore, the determination of whether the Bucky-diamond structure for a $Si_nC_m$ ($m+n=147$) cluster exists must rely on an extensive search of all plausible structural configurations of $Si_nC_m$ clusters. Even though the size of the individual cluster ($m+n=147$) is well within the capability of the DFT-based methods, the large number of plausible compositions of the cluster still makes the search too expensive for the DFT-based calculations. Hence we chose to use the simulation scheme based on the SCED-LCAO approach to conduct the search.

**III. Towards a Search for the Bucky Diamond $Si_nC_m$ Structure: Tuning the Composition and Si/C Arrangements in the Network Created from *3C*-SiC**

The search for the $Si_nC_m$ ($m+n=147$) Bucky-diamond structure was conducted using the initial configurations of the clusters of various compositions based on the following scenarios. (1) The initial configurations of the $Si_nC_m$ clusters were constructed from a spherically truncated *3C*-SiC diamond-like network for $m+n=147$ (containing 11 completed shells) with atoms arranged as either Si-core/C-shell, or Si-shell/C-core, or segregated Si/C configurations of various plausible $m/n$ combinations, using the bulk Si-C equilibrium bond length (1.97 Å) for their nearest neighbor distances, or (2) the initial configurations were constructed by truncating the bulk *3C*-SiC according to the bulk *3C* arrangement of Si/C atoms, namely one shell of Si atoms followed by another shell of C atoms and *etc*. until $m+n=147$ with either Si or C atom at the center (first shell), using the nearest neighbor C-Si equilibrium bond length. The results showing the relaxed configurations from some of the more relevant initial configurations are given in Fig. 2.

From Fig. 2, it is seen that the Bucky-diamond structure $Si_nC_m$ emerges for the magic composition corresponding to $n=68$ and $m=79$. It is rewarding that the search resulted in the discovery of a stable $Si_{68}C_{79}$ Bucky-diamond structure, with a 35-atom $Si_{16}C_{19}$ diamond-like core (with the central atom being a C atom) inside a 112-atom $Si_{52}C_{60}$ Fullerene shell, confirming our anticipated scenario that Si and C can form either $sp^2$ or $sp^3$ bonds and these two bonds can co-exist in the same structure. Specifically, for the $Si_{68}C_{79}$ cluster, a $sp^3$-bonded diamond core can co-exist with a $sp^2$-bonded Fullerene to form the Bucky-diamond structure. A detailed analysis of the relaxed Bucky-diamond structure shows that the relaxed 35-atom diamond-like inner core is barely distorted from its unrelaxed counterpart, with only each of its outermost 6 atoms linked to the outer Fullerene shell. The 112-atom $Si_{52}C_{60}$ outer-shell is composed of 12 pentagons and 46 hexagons, satisfying the rule of the fullerene.

The analysis of our search also points to another interesting observation. It should be noted that the specific composition of the cluster $Si_nC_m$, namely $n=68$ and $m=79$, that yields the Bucky-diamond structure, arises from the bulk *3C*-SiC network truncated at $m+n=147$ containing alternating C shells and Si shells. This observation suggests a likely natural route for synthesizing the SiC Bucky-diamond structure.

We have also validated our result for $Si_{68}C_{79}$ Bucky diamond structure by relaxing the 147 atom SiC network created from bulk *3C*-SiC through the DFT based *ab-initio* simulations (VASP) [21]. The calculations were performed within DFT/GGA using plane-wave basis sets, ultra-soft pseudo-potentials (US-PP) [23], and the Perdew and Wang (PW '91) exchange-correlations [24]. We employed the three-dimensional periodic boundary condition with a vacuum region (15 Å) to ensure that there was no interaction between the $Si_nC_m$ clusters. The cut-off energy for the plane wave basis set was 358 eV. The energy convergence for the self-consistent calculation was set to $10^{-4}$ eV, and the structure was relaxed using the conjugate-gradient (CG) algorithm until the atomic forces were less than $10^{-3}$ eV/Å. The comparison between the two calculations is shown in Fig. 3. It can be seen that the two calculations yielded almost identical results when started from the same identical initial configuration (*i.e.,* the truncated bulk *3C*-$Si_{68}C_{79}$ network).

## IV. The Mechanism Underpinning the Formation of the Bucky-Diamond Structure

The initial configuration of the $Si_nC_m$ cluster corresponding to the atomic arrangement of bulk *3C*-SiC network truncated at 147 atoms with a C atom at the center has a composition $Si_{68}C_{79}$. It is composed of 11 completed shells with alternating C-occupied and Si-occupied shell. The numbers of atoms in shells from shell 1 through shell 11 are 1, 4, 12, 12, 6, 12, 24, 16, 12, 24, and 24 respectively. Thus the outer shell of the interior core (*i.e.* 35-atoms occupying the first 5 shells) has only 6 carbon atoms while the outermost shell of the remaining 6-shell 112-atom configuration has 24 carbon atoms.

The initial configuration is depicted in Fig. 4a. It can be seen that there are groups of two adjacent pairs of unbonded outermost surface (C) atoms (see the surface C atoms (pink balls) with the attached arrows, where the arrowheads indicating the direction of forming C-C bonds). When the configuration is relaxed, both pairs of C atoms will form strong C-C bond to eliminate the dangling bonds. The forming of one pair of strong C-C bond pulls against the respective Si atoms in the $2^{nd}$ outer shell (green balls), each being a nearest neighbor (n.n.) to the C atom (pink balls) on the surface, owing to the strong C-Si bond. This in turn will pull up the C atom in the $5^{th}$ outer shell (red balls), which is the n.n. of both Si atoms in the $2^{nd}$ outer shell (green balls), by the combination of the two Si-C bonds. This series of actions results in the formation of a pentagon on the surface. Likewise, each C carbon (pink balls) of the other pair of induced C-C bond formed on the surface pulls against its own respective n.n. Si atom on the $2^{nd}$ outer shell (green balls), which in turn leads to the formation of an identical pentagon through the n.n. interaction between the Si atoms and their common neighboring C atom from the $5^{th}$ outer shell (red balls). This pentagon actually faces the previous pentagon associated with the previously discussed (induced) C-C bond. This action also pulls the two Si atoms from the $4^{th}$ outer shell (gray balls) towards their respective neighboring C atom in the bonded pairs, through the Si-C bonds, leading to the formation of a hexagon on the surface adjacent to both pentagons located on opposite sides, with the hexagon having two pairs of C-C bonds facing one another that connect two identical pentagons (Fig. 4b). Finally, a hexagon neighboring the hexagon with two C-C bonds can be seen (Fig. 4b) as formed in the following manner. These two hexagons have a common bond formed between the C atom from the first outer (surface) shell and a Si atom from the $4^{th}$ outer shell (gray balls). As the relaxation process continues, the C-atom in the common C-Si bond that is originally from the first shell pulls its n.n. Si-atom on the $2^{nd}$ outer shell (green balls) towards the surface, which in turn pulls its n.n. C-atom from the $3^{rd}$ outer shell (brown balls). Simultaneously, the Si-atom in the common C-Si bond that is originally from the $4^{th}$ outer shell (gray balls) pulls its n.n. C-atom from the $5^{th}$ outer shell (red balls). This C-atom in turn pulls up its n.n. Si-atom from the $6^{th}$ outer shell (blue balls) towards the C-atom originally on the $3^{rd}$ outer shell (brown balls). The completion of the formation of this hexagon then results from the n.n. interaction between the C atom from the $3^{rd}$ outer shell (brown balls) and the Si from the $6^{th}$ outer shell (blue balls). Thus this formation process involves continuously the atoms from the first through the $6^{th}$ outer shell. The situation is depicted in Fig. 4b. It turns out that this hexagon also neighbors the pentagon as it shares a common C-Si bond. A similar situation gives rise to an identical hexagon which is a neighbor to both the hexagon with two C-C bonds and the pentagon on the opposite side. Together with the two hexagons that already existed, the pentagon on the surface is now surrounded by five hexagons, precisely the way that is necessary for the formation of the Fullerene.

Viewed from the direction perpendicular to the presentation in Fig. 4b, there are three hexagons, each with two opposite C-C bonds. The two opposite C-C bonds connect the hexagon to two pentagons. These three pentagon-hexagon-pentagon ensembles and their environments are completely equivalent. Furthermore, it can be seen from Figs. 4a and 4b that there is an identical pattern, namely, three ensembles of pentagon-hexagon-pentagon with their respective environment, existing at the backside of Fig 4b. In fact, the pattern at the backside is an image of the front side, but rotated by $60°$ about the axis perpendicular to the plane of Fig. 4b and passing through the center of the figure. Thus, there are altogether six ensembles of pentagon-hexagon-pentagon, three at the front side and three at the backside. Specifically, the 12 pentagons in these ensembles occupy the strategic locations that provide the curvature for the formation of the Fullerene shell. This can be seen from Fig. 4c that shows the curving fragment anchored by three neighboring pentagons.

The analysis of the relaxation process of the truncated bulk *3C*-SiC structure with a C-atom at the center that fills the first 11 shells clearly demonstrates how the 112-atom Fullerene shell (composed of 52 Si atoms and 60 C atoms) of the $Si_{68}C_{79}$ Bucky-diamond evolves from the elimination of the dangling bonds of the 24 C-atoms on the outermost (surface) shell. It shows that the elimination of the dangling bonds of the carbon atoms on the surface induces strong C-C bonds between pairs of unbonded C-C atoms on the surface. These strong C-C bonds pull the atoms in subsequent shells towards the surface by n.n. C-Si interactions. This process proceeds until it involves all the atoms in the six filled outermost shells of the truncated *3C*-SiC bulk network. An example that best illustrates this continuous process of incorporating atoms from the first through the $6^{th}$ outer shell to form the 112-atom Fullerene shell is the formation of the hexagon that is the common neighbor of the hexagon with two C-C bonds and its neighboring pentagon as already described previously. It should also be noted that with the exception of the 12 C-C bonds, each in one of the pentagon as required, the rest of the Fullerene network can only be composed of Si-C bonds as the formation of the network is driven by n.n. Si-C interactions. This is indeed the case exhibited by the resulting Fullerene.

The next shell after the $6^{th}$ outer shell of the initial configuration of the $Si_nC_m$ cluster with a C-atom at the center obtained by truncating the bulk *3C* network at $n+m=147$ (*i.e.*, $n=68$ and $m=79$) is only occupied by six C atoms. When the equilibrium configuration of the $Si_{68}C_{79}$ Bucky-diamond cluster is established, our analysis shows that the direct distances between the 6 surface atoms of the 35-atom inner core and atoms on the 112-atom Fullerene shell range from 1.97 to 4.28 Å. This then indicates that the inner core is only weakly bonded to the outer Fullerene shell, suggesting that the 35-atom inner core will maintain more or less the unperturbed diamond-like *3C*-SiC configuration. Hence the resulting equilibrated $Si_{68}C_{79}$ Bucky-diamond structure is composed of a 35-atom diamond-like core weakly suspended in the cage of a 112-atom Fullerene shell, a concrete example of the co-existence of the $sp^2$ and $sp^3$ bonding in the same structure.

**V. The Relative Stability of the Bucky-diamond $Si_{68}C_{79}$ Structure with Respect to the Cage Structure**

In our study on all families of C clusters, we have shown that although the Bucky-diamond structure is more stable compared to its corresponding compact cluster of the same size, it is less stable compared to the cage structure of the same size [17]. We would like to ascertain whether this scenario is still valid in the case of SiC-based clusters. Since the SiC Bucky-diamond structure results from the relaxation of the truncated *3C*-$Si_{68}C_{79}$ network, it apparently is the more stable structure among compact clusters of the same composition (*e.g.*, $Si_{68}C_{79}$ clusters in the $4^{th}$ row of Fig. 2). Therefore, we focus our investigation on the relative stability of the SiC Bucky-diamond structure with respect to the cage structure of the same composition.

Previously, we have studied the possible conversion of a C Bucky-diamond structure to its corresponding cage structure by finite temperature MD simulation scheme based on the SCED-LCAO Hamiltonian [17, 18]. It resulted in first the disintegration of the interior diamond core as the carbon Bucky-diamond was heated and then the migration of the interior atoms to the surface. Hence we believe that, in the case of the SiC Bucky-diamond, the inner diamond core will also disintegrate first when the SiC Bucky-diamond is heated.

In this study, we heated the SiC Bucky-diamond from 0 K. At T=1800 K, the inner 35-atom diamond core started to disintegrate. The atoms that broke off from the inner core started to migrate towards the Fullerene shell at T=2000 K. This process continued until most of the interior atoms reached the surface. At T=2600 K, there was hardly any interior atom left. When the resulting structure was slowly quenched to 0 K, a reasonably smooth cage structure at a lower binding energy (-7.881 eV/atom) compared to that of the Bucky-diamond structure (-7.610 eV/atom) finally emerged, as can be seen in Fig. 5a. We found that this cage structure is composed of mostly hexagons and pentagons, but also some heptagons and octagons. It is somewhat surprising that the cage structure of the same composition is still more stable compared to the Bucky-diamond structure as the cage structure is dominated by the $sp^2$ C-Si bond while the strength of the $sp^2$ C-Si bond is somewhat weaker than that of the $sp^3$ C-Si bond (7.96 eV/atom for the $sp^2$ C-Si bond vs 8.02 eV/atom for the $sp^3$ C-Si bond). In Fig. 5, the pair distribution functions of the two structures are also shown (Fig. 5b and 5c). It can be seen that there is almost quadrupling of the much stronger C-C $sp^2$ bonds (8.79 eV/atom for the $sp^2$ C-C bond) in the case of the cage structure compared to that of the Bucky-diamond structure, leading to the more stable cage structure. The appearance of other polygons in addition to hexagons and pentagons is indicative that they provide the relief for the strain due to the formation of the cage structure.

## VI. Conclusion

Our search has led to the successful and, to some extent, surprising discovery of the SiC Bucky-diamond structure at the specific composition of $Si_{68}C_{79}$. It confirms our assertion that a pair of Si and C can form a $sp^3$ bond in the bulk and a $sp^2$ bond on the surface of a SiC-based structure. It also provides the evidence that these two types of bonding can co-exist in the same structure. What is tantalizing is that the SiC Bucky-diamond structure actually results from the relaxation of the cluster $Si_nC_m$ obtained by truncating the bulk *3C*-SiC network at $m+n=147$, suggesting a possible route to synthesize the SiC bucky-diamond.

We have also unraveled the mechanism responsible for the formation of the SiC Bucky-diamond structure as due to the induced strong C-C bonds between pairs of un-bonded C-atoms on the surface of the cluster and the ensuing incorporation of all the atoms in the first six outer shells through the nearest-neighbor Si-C interactions. We believe that a similar mechanism must be responsible for the formation of the carbon Bucky-diamond structure while no mechanism for the formation of carbon Bucky-diamond has been explicitly stated in Ref. [25].

The discovery of the stable SiC Bucky-diamond structure confirms similar bonding nature between Si and C atoms and a pair of C atoms, suggesting that SiC-based nano-structures may manifest themselves in forms similar to C-based nano-structures. Our study has highlighted the importance of the composition in the formation of the SiC Bucky-diamond. We are actively investigating the role played by the composition in determining the structures of other families of SiC-based clusters.

**Acknowledgment:** This work was supported by the U.S. DOE (DE-FG02-00ER4582), NSF (DMR-0112824), and KSEF (KSEF: KY Science and Engineering Foundation). This work was conducted in part using the resources of the Cardinal Research Cluster at the University of Louisville, Louisville, KY.

**Figure captions**

Figure 1 (Color online): Total energy/atom versus MD step (1 MD step = 1.5 fs), depicting the relaxations of $Si_{35}C_{112}$ clusters for two different initial configurations: (a) Diamond-like, tetrahedral arrangements for all Si (in the five inner shells) and C atoms (in the six outer shells), (b) 35 Si atoms (corresponding to first five inner shells) in a tetrahedral network and the remaining 112 C atoms arranged in a fullerene-like shell (see the inset, where yellow, gray, and yellow-gray lines represent the Si-Si, C-C, and Si-C bonds, respectively). The optimized structures corresponding to both initial configurations do not lead to Bucky-diamond structures (see, MD step ~ 1900) but to distorted cage-like structures. The inset in Fig.1(a) shows the total and partial pair distribution functions (*i.e.,* $g(r)$ (black), $g_{C-C}(r)$ (red), $g_{C-Si}(r)$ (green), and $g_{Si-Si}(r)$ (blue)) corresponding to the relaxed structure shown in Fig. 1(a). It can be seen that there are no Si-Si bonds but the peaks corresponding to nearest neighbor C-C and C-Si bonds are prominent.

Figure 2 (Color online): Each panel shows initial and relaxed structures of $Si_nC_m$ ($n+m = 147$) for different compositions (as indicated on the first column) and initial configurations (as indicated on the topmost row). Initial configurations considered include: tetrahedral networking of Si and C atom arrangements with *n* Si-atoms in the core and *m* C-atoms on the outer shells*;* *m* C-atoms in the core and *n* Si-atoms on the outer shells; Si and C atoms completely segregated; and finally the cluster cut from *3C-SiC* bulk. The cohesive energies (eV/atom) corresponding to each of the relaxed

structures are also given. Yellow, gray, and mixed yellow-gray lines represent the Si-Si, C-C, and Si-C bonds, respectively.

Figure 3 (Color online): The relaxed Bucky-diamond $Si_{68}C_{79}$ cluster as obtained from the SCED-LCAO-MD method (a) and that using the DFT/VASP (b). Yellow and green balls represent C and Si atoms, respectively.

Figure 4 (Color online): (a) The initial configuration of the $Si_{68}C_{79}$ cluster cut from bulk *3C*-SiC, (b) the relaxed configuration showing a Fullerene shell and a diamond-like core as obtained by SCED-LCAO-MD, and (c) a fragment of the shell showing the structure anchored by three adjacent pentagons. Pairs of C atoms on the $1^{st}$ outer shell (pink balls), indicated by the arrows in (a), form C-C bonds upon relaxation and they become part of the pentagon-hexagon-pentagon ensembles on the surface of the Bucky-diamond cluster. The Pink, green, brown, gray, red, blue, and yellow balls represent the atoms on the $1^{st}$, $2^{nd}$, $3^{rd}$, $4^{th}$, $5^{th}$, and $6^{th}$ outer shells, as well as the inner core, respectively.

Figure 5 (Color online)   The Bucky-diamond $Si_{68}C_{79}$ cluster (left) transforms to a cage-like structure (right) after annealing it to 2600 K and then slowly quenched to 0K. The corresponding cohesive energies (eV/atom) are also given. The pair-distribution functions corresponding to the bucky-diamond (b) and cage structures (c) are also shown.

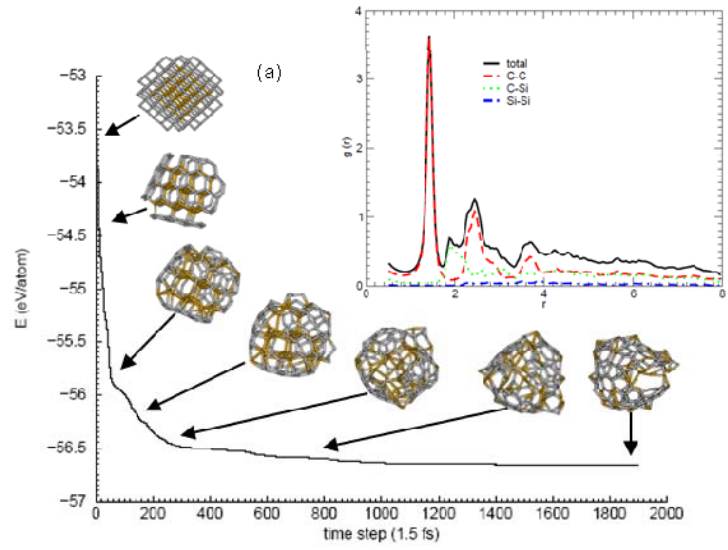

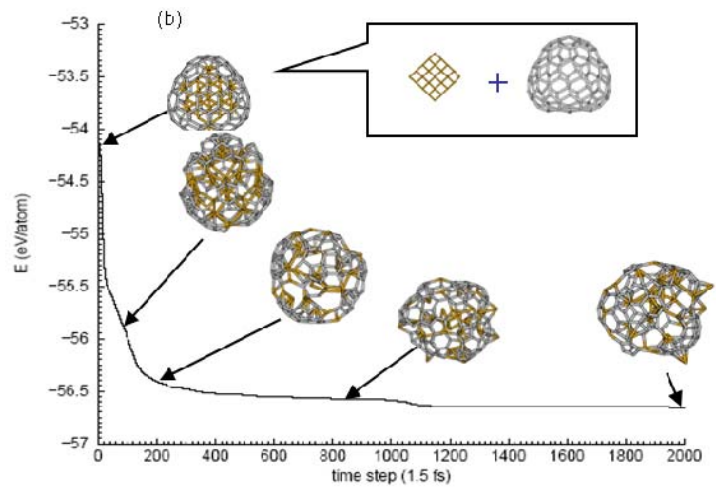

Fig. 1

| $Si_nC_m$ (n + m =1147) | $Si_n$-core/$C_m$-shell | $C_m$-core/$Si_n$-shell | Segregated $Si_nC_m$ Structure | $Si_nC_m$ from 3C-SiC bulk |
|---|---|---|---|---|
| n = 147 | 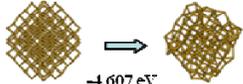 −4.607 eV | | | |
| n = 112 m = 35 | 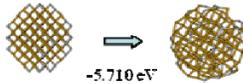 −5.710 eV | 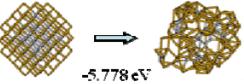 −5.778 eV | 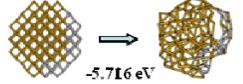 −5.716 eV | |
| n = 79 m = 68 | 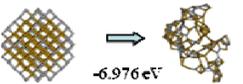 −6.976 eV | 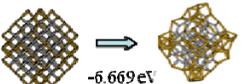 −6.669 eV | 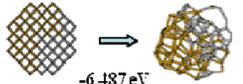 −6.487 eV | 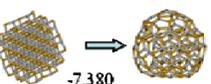 −7.380 eV |
| n = 73 m = 74 | 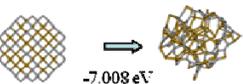 −7.008 eV | 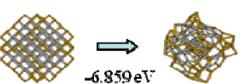 −6.859 eV | 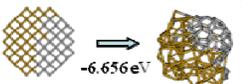 −6.656 eV | |
| n = 68 m = 79 | 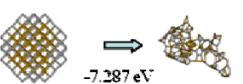 −7.287 eV | 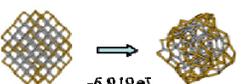 −6.949 eV | 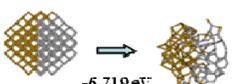 −6.719 eV | 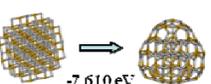 −7.610 eV |
| n = 35 m = 112 | 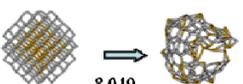 −8.049 eV | 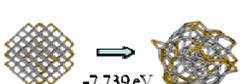 −7.739 eV | 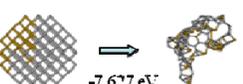 −7.627 eV | |
| m = 147 | | 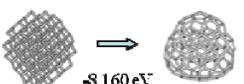 −8.160 eV | | |

Fig. 2

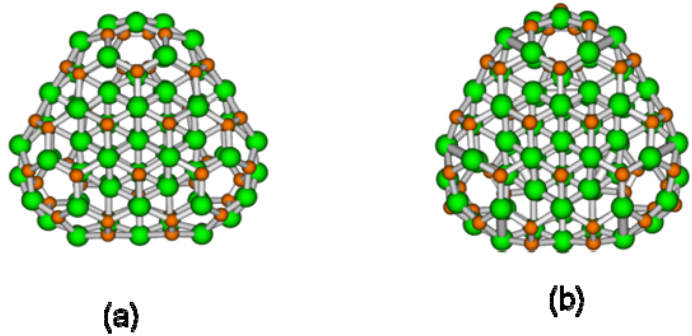

Fig. 3

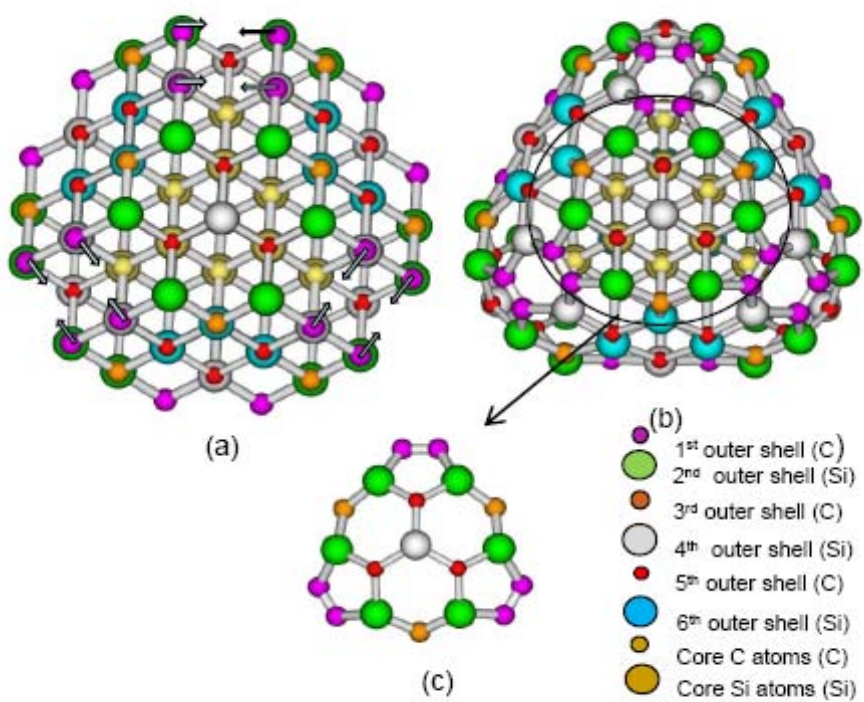

Fig. 4

Fig. 5